\documentclass[acmlarge,nonacm]{acmart}

\begin{document}

\title[Infrastructuring Pop-Up Cities with “Social Layer”]{Infrastructuring Pop-Up Cities with “Social Layer”: Designing Serendipitous Co-Livings for Temporary Intentional Communities}

\author{Danwen Ji}
\orcid{}
\affiliation{%
  \institution{Tongji University}
  \city{Shanghai}
  \country{China}
  }
\email{danwen_ji@tongji.edu.cn}

\author{Botao `Amber' Hu}
\orcid{0000-0002-4504-0941}
\affiliation{%
  \institution{University of Oxford}
  \city{Oxford}
  \country{UK}
  }
\email{botao.hu@cs.ox.ac.uk}

\begin{abstract}
After the pandemic, a new form of "pop-up city" has emerged—co-living gatherings of 100–200 people for 4–8 weeks that differ from conferences and hack houses. These temporary intentional communities leverages existing urban infrastructure, blending daily life (housing, meals, care) with self-organized activities like learning, creating, and socializing. They coordinate bottom-up programming through an "unconference" system for identity, calendaring, RSVP, and social discovery that fosters spontaneous, serendipitous, enduring ties. This paper examines the design of "Social Layer," an unconferencing system for pop-up cities. We studied its real-world deployment in ShanHaiWoo (Jilin, China, 2023), muChiangmai (Chiangmai, Thailand, 2023), Edge Esmeralda, Edge Esmeralda (Healdsburg, CA,USA, 2024), Aleph (Buenos Aires, Argentina, 2024), and Gathering of Tribe (Lisbon, Portugal, 2024). Our findings distill: (1) the strong concept "scaffolded spontaneity"—infrastructural affordances that balance structure with openness, amplifying participant agency while maintaining privacy and lightweight governance; (2) design implications for design researchers working on pop-up cities. 
\end{abstract}


\begin{teaserfigure}
    \centering
    \includegraphics[width=1\linewidth]{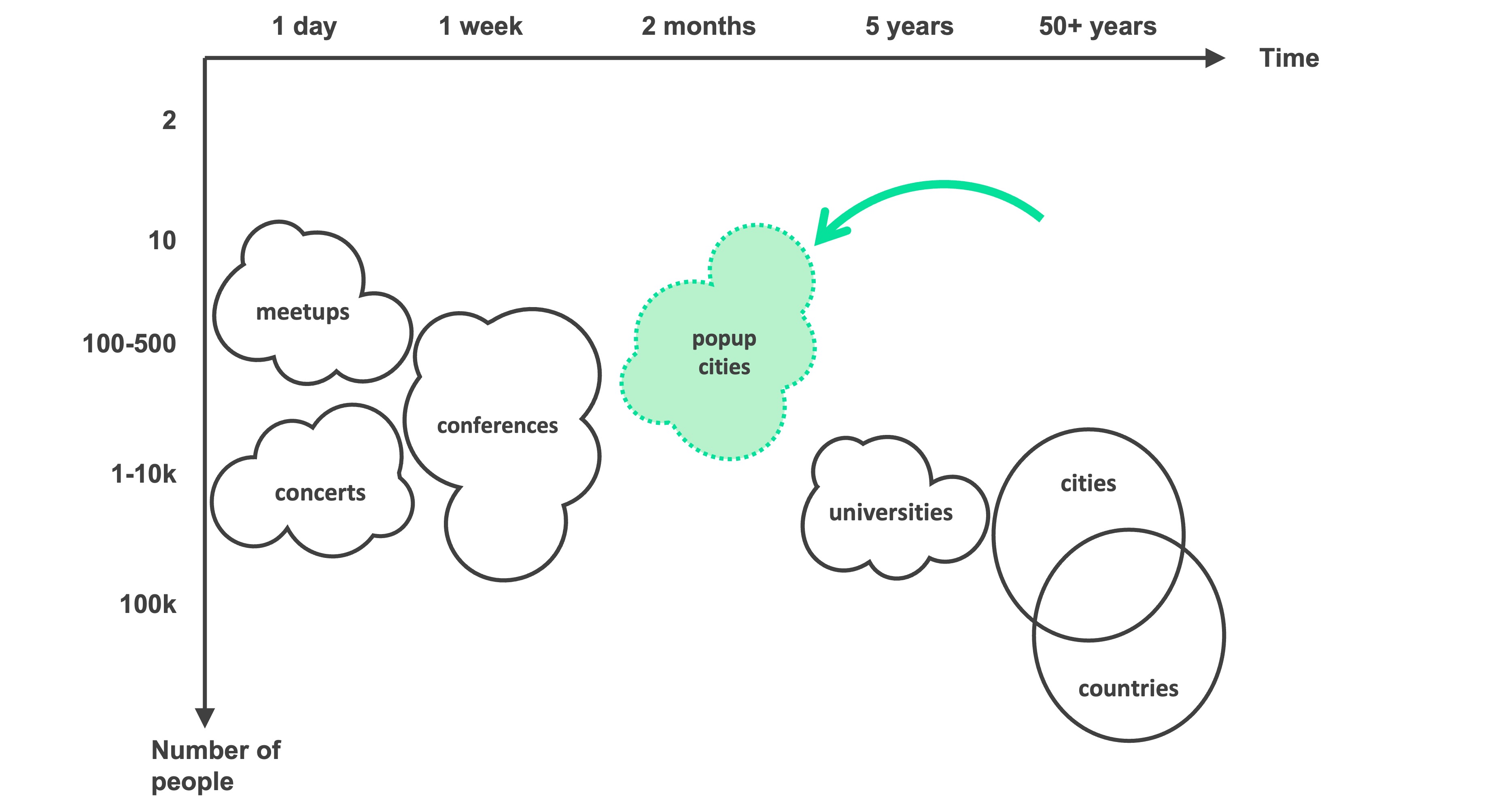}
    \caption{The Missing Piece: Coordinating Sustainable Communities. 
    Adapted and redrawn from a slide by Janine Leger (2023), originally credited to Balaji Srinivasan.}
    \label{fig:teaser}
\end{teaserfigure}

\maketitle

\section{Introduction}

For most of modern history, sustained co‑living has meant marriage and family life, student residencies in universities, or long‑term inhabitation of cities. Shorter formats—workshops, festivals, conferences—create bursts of co‑presence and exchange but rarely support the texture of everyday life together. After pandemic, emerging between these poles is a form of collaborative living that is longer than a conference yet shorter than a graduate program: "pop‑up cities" \cite{sRisePopupCities2024}—multi‑week, residential, self‑organized, and thematically oriented cohorts that occupy or overlay existing urban infrastructures.  Participants work, learn, and live together while co‑creating an evolving program of events, projects, and rituals.
They function as situated experiments in how strangers become collaborators, how programs take shape without planners, and how communities organize themselves in time-bounded worlds. 

Pop-up cities expose systemic tensions—openness versus safety, spontaneity versus coordination, porous boundaries versus coherent identity—requiring designs that shape relational conditions rather than discrete features. We take “infrastructure” in the CSCW sense—as relational, emergent, and enabling rather than merely technical \cite{starStepsEcologyInfrastructure1994,Star1999Ethnographya,starHowInfrastructure2010,karastiInfrastructuringParticipatoryDesign2014,ledantecInfrastructuringFormationPublics2013,hillgrenPrototypingInfrastructuringDesign2011a}. A socio‑technical infrastructure for pop‑up cities must make participation visible and legible  \cite{ericksonSocialTranslucenceApproach2000}), lower activation energy for hosting and attending, protect privacy while establishing trust, and allow communities to “breathe”—tightening to converge, loosening to explore. Designing for pop-up cities also requires attending to emergent social networks \cite{ahrensQualitativeNetworkAnalysis2018}, understanding how infrastructures shape opportunities \cite{murphySystemicTheoriesChange2021,murphyLeverageAnalysisMethod2020}, and treating friction as generative rather than eliminable \cite{ozkaramanliDilemmasConflictsSystemic2022}. Serendipity research further highlights the roles of affordances, discovery, and structured openness \cite{makriComingInformationSerendipitously2012,buschTheorySerendipitySystematic2024,bjornebornThreeKeyAffordances2017,andreDiscoveryNeverChance2009,kotkovSurveySerendipityRecommender2016}, while social network theory emphasizes weak ties, brokerage, and small-world mixing \cite{granovetterStrengthWeakTies1973,burtStructuralHolesGood2004,uzziCollaborationCreativitySmall2005}.

Designing for such temporary intentional communities requires rethinking "program." Conferences rely on detailed schedules, expert-curated tracks, and tightly managed keynote economies. Degree programs build semester arcs, syllabi, assessments, and credentialing. Both are heavyweight, predictable, and optimized for coordination under certainty. By contrast, pop-up cities aim for what participants often describe as "serendipitous criticality": enough structure to build trust, safety, momentum, and activation; enough openness to enable discovery, self-direction, and serendipity. This aligns with systemic design principles emphasizing emergent behaviors, and evolutionary change \cite{vanderbijl-brouwerSystemicDesignPrinciples2020c, vanderbijl-brouwerSystemicDesignReasoning2024}. These communities blend the weak structures of unconferences \cite{Park2023Future} with the everydayness of co-living. This raises a core question for HCI and design research: 

\emph{How can we design infrastructures for serendipity at the scale of a temporary city, and what key characteristics support it?}

This paper reports our findings through a Research-Through-Design \cite{zimmermanResearchDesignHCI2014} process involving continuous design and deployment of "\textit{Social Layer}"\footnote{\url{https://www.sociallayer.im/}}, a modular online infrastructure supporting serendipitous co-living in pop-up cities. It scaffolds spontaneity within temporary intentional communities and reconfigures relational conditions through privacy-preserving credentials for identity and trust, lightweight event initiation and RSVP mechanics, visibility cues, and tools for social discovery and portable community memory. We distill our knowledge through deployments in real-world communities—as "living laboratories" \cite{pipekInfrastructuringIntegratedPerspective2009,bjorgvinssonParticipatoryDesignDemocratizing2010d,karastiInfrastructuringParticipatoryDesign2014}.

We contribute:
(1) an real-world exploration of pop-up city infrastructure as a systemic design space within temporary urbanism and intentional community \cite{bishopTemporaryCity2012,firthUtopianismIntentionalCommunities2019,sagerPlanningIntentionalCommunities2018a,quinnUnderstandingUrbanCultural2025,meijeringIntentionalCommunitiesRural2007};
(2) an articulation of Social Layer as an infrastructural intervention that modulates emergent serendipitous co-living in pop-up cities; and
(3) design implications centered on the strong concept of "scaffolded spontaneity"—operating near the edge of chaos \cite{kauffmanOriginsOrderSelfOrganization1993,hollandHiddenOrderHow2003}—which supports serendipity, trust, plurality, and collective agency in time-bounded worlds.

\section{Background}

\subsection{Intentional Communities, Temporary Urbanism, and Pop-up Cities}

Human experiments in intentional community have a long history. From ecovillages and artist colonies to 1970s utopian communes and contemporary co-living startups, such initiatives explore alternative forms of collective living, governance, and social innovation \cite{brownIntentionalCommunityAnthropological2002,shenkerIntentionalCommunitiesRoutledge2011}. They often reconfigure daily life—housing, work, leisure, decision-making—around shared values. While many communities remain small or persistent, recent years have seen the rise of temporary intentional communities, bringing large groups together for bounded periods. Some evolve into long-term formations, echoing historical precedents such as Chautauqua\footnote{https://www.chq.org/}, demonstrating potential for sustained social experimentation.

Urban studies frame “Temporary Urbanism” as tactical, cultural, and strategic interventions that reconfigure space and social relations \cite{schusterEphemeraTemporaryUrbanism2001,stevensTemporaryTacticalUrbanism2023,mouldTacticalUrbanismNew2014,bishopTemporaryCity2012,firthUtopianismIntentionalCommunities2019,andresAdaptableCitiesTemporary2025}. Festivals, pop-ups, and residencies mobilize existing urban infrastructures while layering temporary civic order. Burning Man illustrates temporary cities as socio-cultural infrastructures shaping identities, practices, and collaborative production \cite{turnerBurningManGoogle2009,chenArtisticProsumptionCocreative2012,smithFestivalsCityContested2022}.

Pop-up cities inherit this lineage but extend it toward hybrid work–learning–living arrangements, often among digitally networked residents. Zuzalu\footnote{https://wiki.p2pfoundation.net/Zuzalu} assembles a temporary, invitation-based community to co-live and co-work around technology, governance, and decentralized innovation, functioning as a living lab for prototyping social forms, governance models, and collaborative practices \cite{fitzpatrickMethodologicalPluralismPractice,vanderbijl-brouwerSystemicDesignPrinciples2020c,perenoTeachingSystemicDesign2025,landa-avilaRethinkingDesignComplex2022}. Inspired by Zuzalu, similar pop-up cities have emerged globally\footnote{https://www.nsforum.net/posts/a-recap-of-zuzalu-inspired-pop-up-cities}, using temporary, mission-driven settlements to build intentional communities and test ideas that typically circulate only online. These arrangements foreground emergent networks, relational infrastructures, improvisation at scale, and methodologically pluralistic approaches, reflecting systemic design’s emphasis on multiple epistemologies and reflective practice \cite{fitzpatrickMethodologicalPluralismPractice, vanderbijl-brouwerSystemicDesignReasoning2024}.

\subsection{Serendipity and Groupware}

Serendipity—positive, meaningful, and unexpected discovery—has been studied across information science, HCI, and organizational scholarship \cite{copelandSerendipityScienceDiscovery2019,mccay-peetMeasuringDimensionsSerendipity2011,andreDiscoveryNeverChance2009,makriComingInformationSerendipitously2012,buschTheorySerendipitySystematic2024}. Rather than discrete events, serendipitous experiences emerge relationally under exposure, novelty, and interpretive readiness. Computational systems operationalize serendipity via novelty, diversity, and coverage \cite{geAccuracyEvaluatingRecommender2010,kotkovSurveySerendipityRecommender2016}, increasing the probability of meaningful encounters without over-determining interactions.

Organizational scholarship connects weak-tie bridging and network range to creativity and innovation \cite{granovetterStrengthWeakTies1973,burtStructuralHolesGood2004,uzziCollaborationCreativitySmall2005,tortorielloBridgingKnowledgeGap2012}. These studies show that individuals embedded in diverse and loosely coupled networks are more likely to access non-redundant information, encounter divergent perspectives, and form novel collaborations \cite{frydenbergCultivatingSerendipityDesign}. Urban research on “third places” \cite{oldenburgOurVanishingThird} and social infrastructure \cite{klinenbergPalacesPeopleHow2018} shows informal environments fostering micro-encounters and broader social cohesion. Designing for serendipity entails cultivating affordances, enhancing visibility and legibility (social translucence), and enabling gentle temporal rhythms \cite{bjornebornThreeKeyAffordances2017,buschPlannedLuckHow2021a,ericksonSocialTranslucenceApproach2000}. In pop-up cities, these principles intersect with systemic design methods, as participatory and embodied practices structure emergent learning, collaboration, and reflection \cite{fitzpatrickMethodologicalPluralismPractice}.

\subsection{Infrastructuring for Unconferencing, Events, and Temporary Co-living}

“Infrastructure” in CSCW and STS is relational and becomes visible through breakdowns or design interventions \cite{Star1999Ethnographya,starStepsEcologyInfrastructure1994}. Infrastructuring foregrounds ongoing socio-material work and coordination across heterogeneous actors, often via boundary objects \cite{pipekInfrastructuringIntegratedPerspective2009,karastiInfrastructuringParticipatoryDesign2014,bjorgvinssonParticipatoryDesignDemocratizing2010d,StructureIllStructuredSolutions1989,bowkerSortingThingsOut1999}. Contemporary platforms blur lines between application and infrastructure \cite{plantinInfrastructureStudiesMeet2018,plantinReintegratingScholarlyInfrastructure2018}.

Pop-up cities layer municipal and digital infrastructures, combining Open Space Technology principles with everyday logistics \cite{owenOpenSpaceTechnology2008}. OST self-organizing mechanisms structure emergent participation while maintaining coherence. Social Layer \footnote{https://app.sola.day} operationalizes this stance digitally, supporting the continuous surfacing of invitations, RSVP, and gentle coordination. Trust and privacy are supported through identity systems enabling “proofs without disclosure” (e.g., zero-knowledge proofs, Zupass\footnote{\url{https://github.com/proofcarryingdata/zupass}}) and proof-of-attendance artifacts (POAPs\footnote{\url{https://poap.xyz}}) that record participation without centralizing identity.

These arrangements exemplify how systemic design scaffolds emergent community practices, balances openness and safety, and integrates infrastructure, rhythms, and technology into cohesive experimental ecosystems \cite{fitzpatrickMethodologicalPluralismPractice,vanderbijl-brouwerSystemicDesignPrinciples2020c,perenoTeachingSystemicDesign2025,vanderbijl-brouwerSystemicDesignReasoning2024,gossDesignCapabilityWhen2024b,aulisioSystemicDesignSustainable2024}. These studies highlight the methodological pluralism of systemic design, showing how multiple approaches, reflective practices, and context-sensitive strategies inform the design of complex communities.

\section{Method}

We follow a research‑through‑design (RtD) approach \cite{zimmermanResearchDesignHCI2014,zimmermanAnalysisCritiqueResearch2010,gaverWhatShouldWe2012a,zimmermanResearchDesignMethod2007,bowersLogicAnnotatedPortfolios2012}, analyzing the Social Layer artifact in use across multiple deployments while iteratively refining its architecture to investigate how infrastructural design can scaffold serendipity in temporary intentional communities. 

Rather than formulating a fixed hypothesis and running controlled experiments, we:
\begin{itemize}
\item	Co‑designed and implemented Social Layer with organizers of pop‑up cities;
\item	Deployed and adapted the system across multiple \item	Analysed how infrastructural choices shaped participation, serendipity, and memory.  
\end{itemize}

The authors were embedded as designers, facilitators, and participants in several of the deployments. This insider–outsider position afforded rich access to configuration decisions, breakdowns, and emergent practices, while also requiring reflexive attention to our own role in shaping the phenomena we study.

Our inquiry integrates three complementary strands of data:
(1) \textbf{Public and archival materials} documenting community operations, programs, and schedules across five deployments (e.g., ShanHaiWoo, muChiangmai, Edge Esmeralda, Aleph, Gathering of Tribe);
(2) \textbf{Design and facilitation field notes} generated through ongoing participation in system integration, onboarding, and community governance; and
(3) \textbf{Platform artifacts and interaction traces}, including configuration schemas, event logs, and interface iterations of the Social Layer system.
Together, these sources provide a multi-scalar perspective—from system architecture and community design choices to participants’ situated improvisations.

We conducted an interpretivist thematic analysis \cite{braunCanUseTA2021}, sensitized by concepts from infrastructuring \cite{karastiInfrastructuringParticipatoryDesign2014,pipekInfrastructuringIntegratedPerspective2009} and social translucence \cite{ericksonSocialTranslucenceApproach2000}. 

We coded the data in three stages: firstly, identifying material and social arrangements that shaped participation (e.g., visibility, modularity, portability); secondly, tracing how these arrangements enabled emergent coordination or trust; and thirdly, abstracting design patterns that constitute the “social layer” as an infrastructural condition for serendipity.

\section{System Design: Solar Layer}

\begin{figure}
    \centering
    \includegraphics[width=1\linewidth]{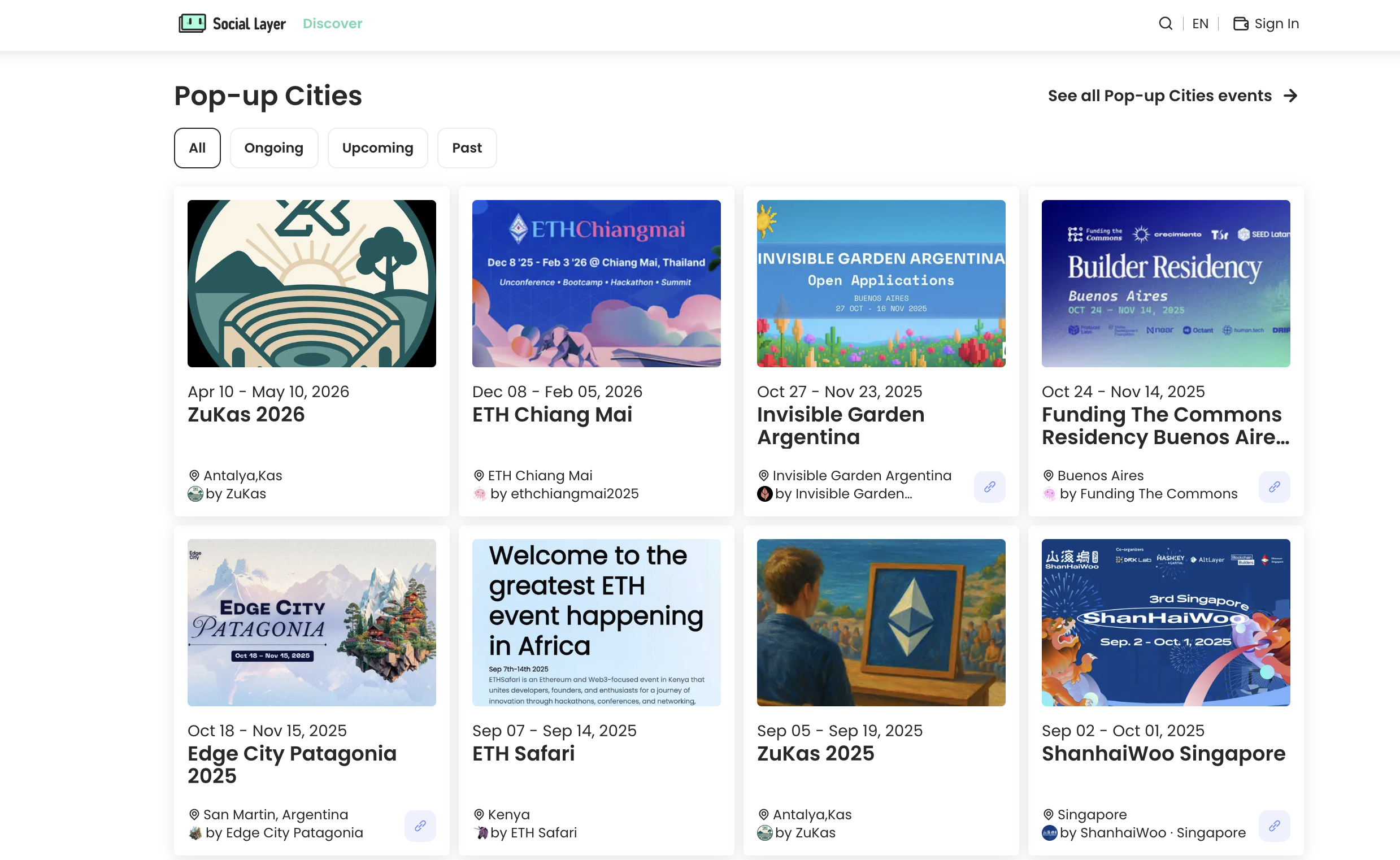}
    \caption{Solar Layer is a modular infrastructure for the design, scale, and sustainability of pop-up communities and cities.}
    \label{fig:sola}
\end{figure}

Social Layer was conceived not as a discrete “app” but as a living infrastructural ensemble co-developed with and within temporal intentional communities. From its earliest prototypes, the system has evolved through cycles of deployment, feedback, and re-composition in contexts such as pop-up cities and residency-style gatherings. Each iteration translated tacit organizer knowledge—how to convene, announce, and host—into concrete interface elements: buttons, forms, and flows that scaffold collective action. In this sense, Social Layer embodies infrastructuring rather than engineering: it is continually shaped by the very communities it supports.

\subsection{Design Principles}

\subsubsection {Identity and Trust}

In temporary or nomadic communities, trust must be established quickly and maintained without relying on heavy credential systems. Social Layer integrates privacy-preserving credentials and lightweight verification cues that allow participants to recognise one another without disclosing unnecessary personal data. Profiles operate less as static self-descriptions and more as “trust surfaces”—showing enough about someone’s presence, history, and contributions to enable accountability without collapsing privacy boundaries.

Different communities articulate different notions of membership and belonging. To accommodate this, Social Layer supports configurable boundary modules: communities can define how identity is verified (for example, invitation tokens, wallet-based proof of participation, peer approval, or open registration). These modular boundaries are not only security mechanisms but also social structures—ways of defining how open or bounded a collective should be. This flexibility enables communities to experiment with new governance models and iterate their own norms of trust.

\subsubsection{Unconference-Style Event Initiation}

The core interaction model re-imagines event creation as an act of invitation rather than administration. Any participant can propose a gathering—an impromptu talk, dinner, or co-working session—by entering minimal information (“what,” “where,” “when”). Events emerge from the periphery of attention rather than from top-down schedules, aligning with the improvisational logic of unconferences and pop-up cities.

The system supports dynamic rescheduling and conflict prevention: participants can adjust times or locations and the platform detects overlapping uses of shared spaces. This reduces the need for dedicated coordinators and releases organisers from heavy logistical labour, allowing them to focus on facilitating content and dialogue instead.

Communities may also define their own rules of structure—for instance, detailed booking policies or spontaneous room allocation—and these can be adjusted over time. Unlike many commercial SaaS systems where rules are hard-coded, Social Layer treats structure itself as an editable layer. This allows communities to collectively negotiate and iterate norms for shared space usage.

\subsubsection {RSVP and Presence Cues}

Attendance in pop-up environments is fluid; people drift between sessions, meals, and projects. Social Layer therefore adopts a lightweight RSVP rather than strict registration. The interface provides affordances for varying degrees of formality as will ; hosts of structured gatherings may activate a simple check-in process with the QR code and the camera in their mobile if needed. And the participant list indicator who is having a mutual topic. 

These cues enact social translucence: they make participation visible enough to coordinate, while preserving the ambiguity necessary for serendipitous encounters. The map interface visualises upcoming or nearby events, supporting low-commitment exploration—participants can discover “what’s happening nearby soon” and spontaneously join.

Instead of algorithmic personalisation, serendipity arises through infrastructural visibility—the system’s spatial and temporal rhythms that let patterns self-organise. This design encourages participants to balance planning with discovery, structure with openness.

\subsubsection {Memory and After-life}

Although the communities using Social Layer are temporary, their interactions produce knowledge, relationships, and artifacts worth carrying forward. The system therefore can be seen as a memory layer that preserves event histories, photos, notes, and participation traces as portable memories. These archived traces form a connective tissue between pop-ups, enabling practices and relationships to travel into future iterations.

This design supports not only co-presence but also after-presence—the lingering infrastructures of connection that sustain collaboration beyond the lifespan of a single residency. Memories are community-owned, exportable, and can seed the setup of future deployments, helping participants build continuity across ephemeral worlds.

\subsection{Architecture and Interfaces}

\subsubsection{Coordinator Dashboard}

The coordinator-facing interface provides configuration and infrastructural management capabilities. From the dashboard, community managers can define the parameters that shape collective coordination (Fig. \ref{fig:dashboard}):

\begin{figure}
    \centering
    \includegraphics[width=1\linewidth]{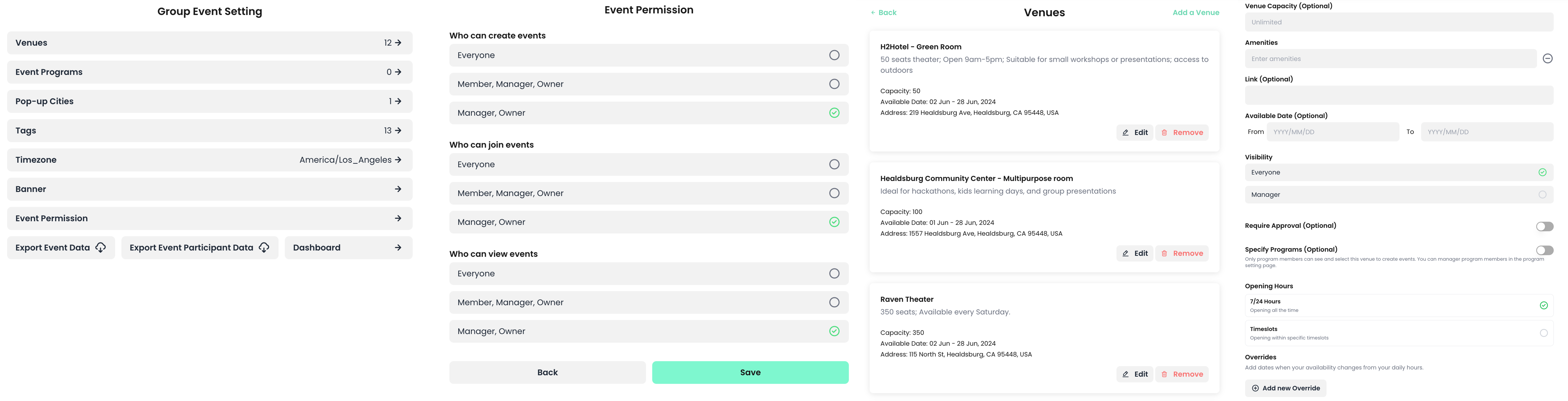}
    \caption{Coordinator Dashboard from Solar Layer}
    \label{fig:dashboard}
\end{figure}

\begin{itemize}
    \item \textbf{Group and Event Settings} — Coordinators can set global parameters such as time zone, event visibility, and participant permissions (e.g., who can create, join, or view events). These settings allow each community to experiment with openness and self-governance.
    \item \textbf{Programs and Venues} — Events are grouped into \emph{Programs} (e.g., thematic residencies) and attached to \emph{Venues} that represent physical or virtual spaces. Each venue includes editable metadata such as location, description, capacity, amenities, availability windows, and opening hours. Coordinators can restrict venues to specific programs or make them globally available.
    \item \textbf{Permission and Role System} — Facilitators, members, and participants can have distinct roles. These roles determine which functions are accessible in a given context—allowing a single user to act as organiser in one residency and as participant in another.
    \item \textbf{Data Export and Transparency} — Event and participation data can be exported, supporting community ownership and further analysis. This functionality treats data not as proprietary but as a shared resource.
\end{itemize}

The dashboard thus constitutes the infrastructural layer of the social fabric—where boundaries, capacities, and rhythms of visibility are collectively defined.

\subsubsection{Participant Interface}

\begin{figure}
    \centering
    \includegraphics[width=1\linewidth]{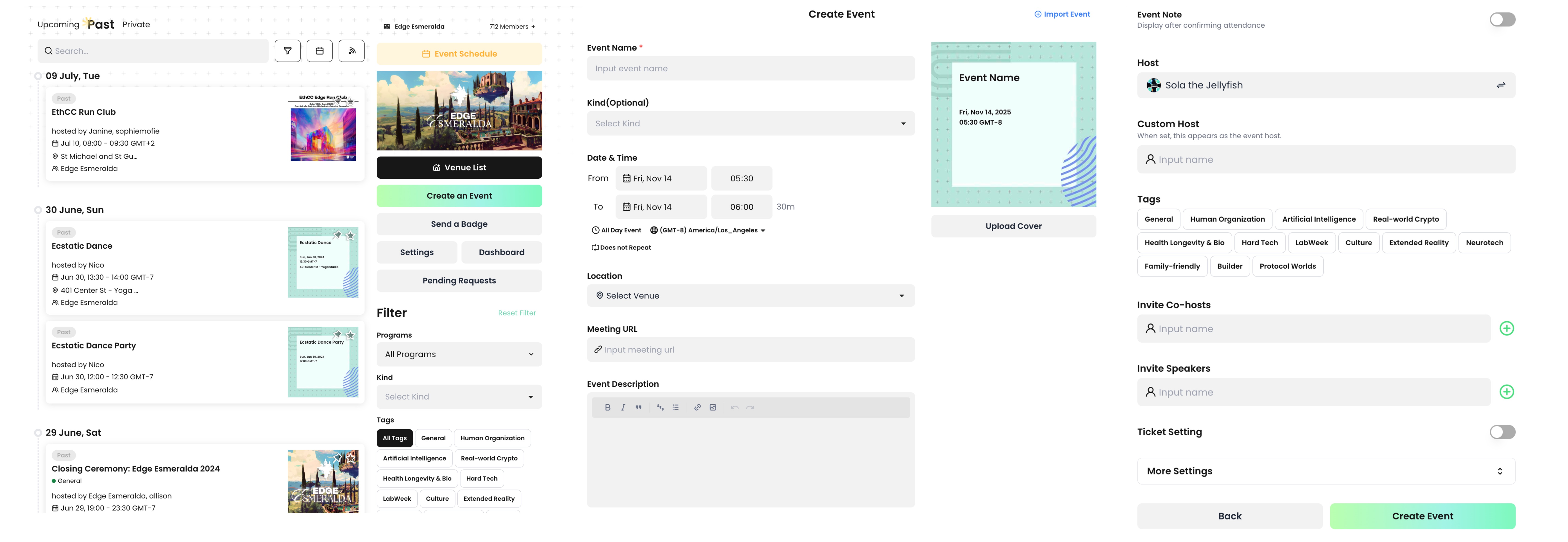}
    \caption{Event Creation Process from Solar Layer}
    \label{fig:eventcreate}
\end{figure}

The participant-facing interface emphasises immediacy, discovery, and serendipity. Its structure supports fluid movement between planned and spontaneous participation (Fig. \ref{fig:eventcreate}).
\begin{itemize}
    \item \textbf{Prominent Event Creation} — A visible \emph{“Create Event”} button anchors the interface. Participants can initiate gatherings with minimal input (title, time, venue), lowering the threshold for contribution. Optional modules allow hosts to add co-hosts, speakers, tickets, or tags, but these remain secondary.
    \item \textbf{Event Schedule and Discovery} — A unified schedule displays all ongoing and upcoming activities across programs. Users can toggle between compact, list, venue, or weekly views. The schedule visualises overlaps and temporal rhythms, supporting coordination through visibility rather than control.
    \item \textbf{Filtering and Navigation} — Events can be filtered by tag, venue, or program, enabling participants to navigate thematic and spatial affinities while still being exposed to adjacent activities. The map view provides a spatial view of ongoing and upcoming events across venues. Each event is geolocated, supporting both navigation and serendipitous encounters.
    \item \textbf{RSVP and Presence Cues} — Participants can mark attendance or interest("star") without committing fully. Hosts may optionally enable QR-based check-in. Presence indicators make participation visible enough for coordination while maintaining the ambiguity conducive to informal joining.
    \item \textbf{Badges and Tickets} — Tickets can be configured as free or paid, with optional qualification badges granting access or privileges. These symbolic elements support informal reputation and reciprocity within the community.
\end{itemize}

Together, these features create a digital topology where structured programs coexist with improvisational events, allowing planned and serendipitous gatherings to emerge side by side.

\subsubsection{Shared Schedule and Spatial Coordination}

\begin{figure}
    \centering
    \includegraphics[width=1\linewidth]{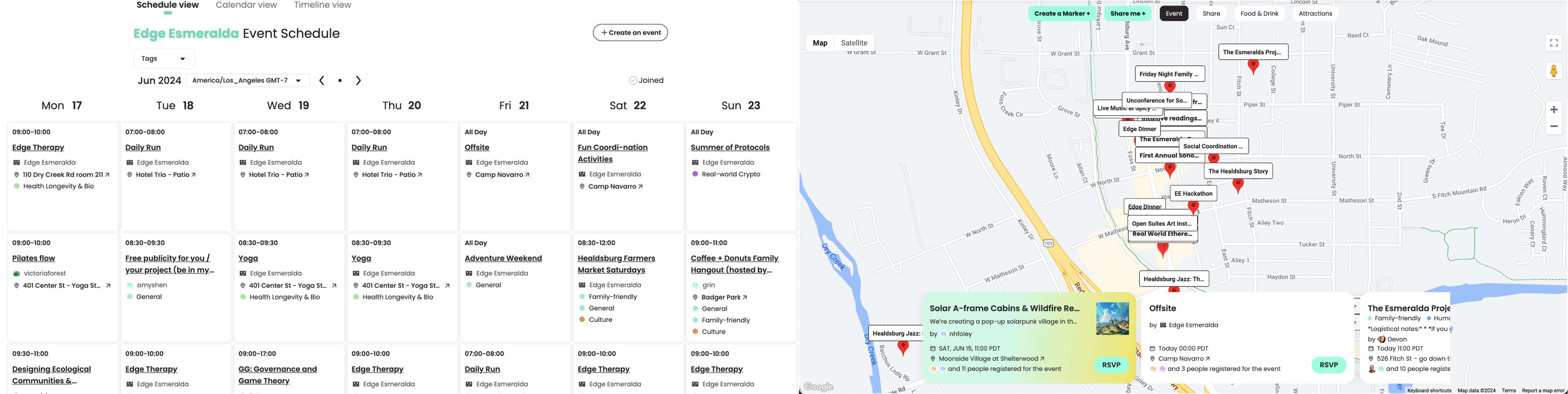}
    \caption{Shared Schedule and Map View from Solar Layer}
    \label{fig:schedule}
\end{figure}

The shared schedule acts as the system’s central coordination surface. Each event is embedded within time and place, connecting the digital interface to the lived geography of the pop-up city or residency. The schedule displays daily and weekly rhythms—yoga sessions, deep-work sprints, talks, dinners—and makes them legible to all participants (Fig. \ref{fig:schedule}).

By visualising temporal proximity and venue co-location, the schedule surfaces potential intersections: users might notice overlapping sessions or nearby gatherings and decide to join spontaneously. This infrastructural visibility replaces algorithmic recommendation with a collectively perceivable temporal map.

\subsubsection{Configurability and Portability}

All data and configurations are exportable for community member with the permissions, allowing communities to migrate, fork, or recompose the system for future residencies. The architecture treats each deployment as a version rather than an instance: communities can inherit prior structures while reshaping them to local conditions. This supports continuity across temporary gatherings and builds an evolving ecology of interconnected infrastructures.
  
\section{Results: Diving into the Co-Livings of Pop‑Up Cities with Social Layer}
Through a research-through-design process, we examined how infrastructural design might scaffold serendipity at the scale of temporary, intentional communities. Social Layer evolved from a speculative prototype into a functioning socio-technical infrastructure adopted by multiple pop-up residencies. Each deployment tested how lightweight coordination, visibility, and modular trust mechanisms could support the emergence of improvisational order. Each deployment not only tested design hypotheses about serendipity and participation but also enacted infrastructuring in practice—embedding the system into local contexts, governance patterns, and event ecologies.

\subsection{Macro-level Observations Across Communities}

Across its deployments, Social Layer has supported a diverse network of pop-up residencies and intentional gatherings, including \emph{Shanhaiwoo 2023}\footnote{https://www.shanhaiwoo.com/recap}, \emph{muChiangmai 2023}\footnote{https://the-mu.xyz/blog/muchiangmai}, \emph{Edge Esmeralda 2024}\footnote{https://www.edgeesmeralda.com/2024}, \emph{Aleph 2024} \footnote{https://crecimiento.build/blog/aleph}, and \emph{Gathering of Tribe 2024}\footnote{https://the-gathering.earth/}. These deployments ranged from week-long micro-residencies to month-long community experiments, with participant communities ranging from roughly 100 to 800 members. Collectively, they served as living laboratories for infrastructuring temporary social worlds---spaces where digital coordination tools, physical proximity, and shared intent intersected to create ephemeral but socially dense collectives.

We summarize macro-level participation metrics across these deployments in Table \ref{statistics}. 

\begin{table}[htbp]
\centering
\caption{Statistics Across Deployments}
\label{statistics}
\begin{tabular}{lrrrrrr}
\toprule
\textbf{Comm.} & 
\textbf{Dur. (d)} & 
\textbf{Evts} & 
\textbf{Self-Org.} & 
\textbf{Hosts} & 
\textbf{Parts.} & 
\textbf{Part.-Instances} \\
\midrule
Shanhaiwoo & 27 & 133 & 126 & 40  & 102 & 915  \\
muChiangmai & 47 & 118 & 46 & 25 & 120 & 571 \\
Edge Esmeralda & 28 & 554 & 405 & 136 & 582 & 4952 \\
Aleph & 28 & 417 & 173 & 86  & 889 & 4496 \\
Gathering of Tribe & 6 & 364 & 356 & 89  & 151 & 1100 \\
\bottomrule
\end{tabular}
\vspace{2mm}
\begin{minipage}{0.9\linewidth}
\footnotesize
\textit{Note.} 
\textbf{Comm.} = name of the pop-up city; 
\textbf{Dur. (d)} = duration in days; 
\textbf{Evts} = total number of events; 
\textbf{Self-Org.} = self-organized (member-initiated) events not hosted by official organizers; 
\textbf{Hosts} = number of unique hosts; 
\textbf{Parts.} = number of unique participants; 
\textbf{Part.-Instances} = total participation instances (including multiple attendances by the same individual). 
Only recorded RSVP data are included; many participants joined directly without registration.

\end{minipage}
\end{table}

Across all deployments, participants collectively hosted 1,586 events, of which 1,106 (69.7\%) were self-organized after the community had already begun. This high proportion indicates that spontaneous, member-driven activity was not peripheral but central to how these temporary collectives evolved. As residents became familiar with shared rhythms and infrastructures, they increasingly initiated their own gatherings, transforming the platform from a coordination tool into a medium for emergent authorship.

The intensity and form of self-organization varied considerably across sites. 
\emph{Gathering of Tribe} demonstrated an almost fully emergent mode, with 98\% of its 364 events organized within a six-day period—an ecosystem of serendipitous encounters arising from high temporal density. 
\emph{Shanhaiwoo} similarly exhibited strong “bottom-up” dynamism (95\%), where informal dinners, discussions, and workshops continuously reconfigured social connections. 
By contrast, \emph{muChiangmai} and \emph{Aleph} balanced structured programming with open-ended extensions (41\%), blending planned coordination with unplanned improvisation. 
\emph{Edge Esmeralda}, showing how even large-scale communities can sustain distributed, serendipity-driven creation once infrastructural conditions are in place.

Across these contexts, serendipity operated less as random chance and more as an infrastructured affordance—a pattern of social emergence enabled by lightweight tools for visibility, coordination, and invitation. The Social Layer thus facilitated not merely participation but the continuous re-authoring of community life through unplanned yet consequential encounters.

\subsection{Micro-Level Observations: In Case of ShanHaiWoo 2023}

\begin{figure}
    \centering
    \includegraphics[width=0.75\linewidth]{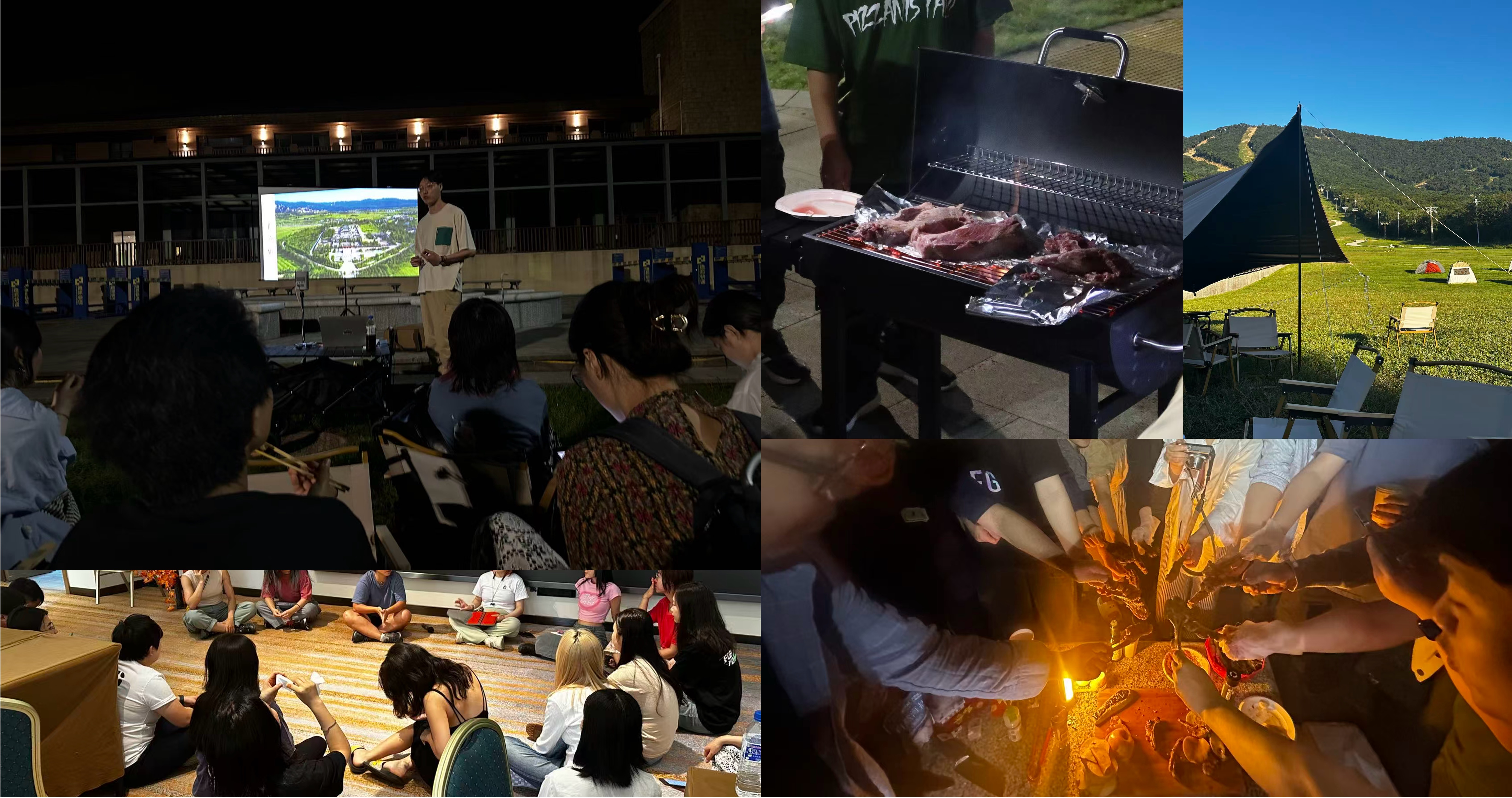}
    \caption{Moments in ShanHaiWoo}
    \label{fig:shanhaiwoo}
\end{figure}

Zooming in, \textit{ShanHaiWoo 2023} offers a rich view of micro-level affordances and social emergence.
Held in Beidahu, Jilin, over four weeks, the residency gathered around 100 participants from art, design, technology, and regenerative culture.
Over 120 events were recorded—many spontaneous, self-initiated, and socially contagious through the platform’s one-click event creation.(Fig. \ref{fig:shanhaiwoo}).

\begin{quote}
\textit{“Social Layer made it effortless to start something—it removed permission, roles, and the need for approval.”(P1)}\\
\textit{“It was a social-technical system that amplified individual agency—turning a personal spark into collective action.”(P2)}\\
\textit{“It’s not the tool that created results—it simply set the condition for serendipity out of chaos.”(P3)}
\end{quote}

From the analysis of field notes and participant reflections, five intertwined patterns emerged.

\subsubsection{Emergent Self-Organization and Collective Rhythm}

Participants described a dynamic equilibrium between freedom and coherence—\textit{a boundary that felt both free and safe}.
Through minimal scaffolding (shared calendar, chat channel, physical notice board), self-organization flourished:

\begin{quote}
\textit{“After the first unconference, spontaneous events filled every hour until everyone was exhausted.” (P5)}\\
\textit{“People from different disciplines came together and self-organized activities—there was never a shortage of ideas.” (P4)}
\end{quote}

The infrastructural visibility of others’ actions created temporal rhythm: morning yoga, afternoon unconferences, late-night bonfires—all loosely synchronized yet emergent.

\subsubsection{Playful Improvisation and Situated Making}

Spontaneous play and experimentation became social bonding rituals.

\begin{quote}
\textit{“A group of not-so-outdoorsy people spent an hour putting up four tents, held the first ‘mountain assembly,’ and welcomed newcomers through a face-to-face onboarding ritual.” (P7)}\\
\textit{“While playing around, we learned new things, and while learning, new ideas emerged.” (P9)}
\end{quote}

Improvisation was central—not as deviation but as situated making: acting in context with others to shape shared meanings and material conditions.

\subsubsection{Amplified Agency and UGC-Style Participation}

\begin{quote}
\textit{“Within three days we launched a Women Coding Camp—friends who were just chatting the day before suddenly became mentors.” (P15)}
\end{quote}

The system’s low-friction creation mechanism transformed passive attendance into user-generated culture (UGC).
Participants internalized a permissionless ethos: “anyone can host,” leading to a cascade of peer-driven initiatives that blurred the line between organizer and participant.

\subsubsection{Real Encounters and Social Trust}

\begin{quote}
\textit{“On a hike, we could trace an idea from its spark to realization—through intimate, one-to-one exchanges.” (P12)}\\
\textit{“Shared meals became the seed for an on-chain art game.” (P11)}
\end{quote}

Physical co-presence intertwined with digital visibility, fostering trust through encounter.
Serendipitous proximity—seeing others’ open events and joining spontaneously—enabled deeper relational bonds than planned networking could achieve.

\subsubsection{World-Building and Collective Imagination}

\begin{quote}
\textit{“In just a few days, the snowfield turned into a miniature world—a place of collisions and surprises.” (P16)}\\
\textit{“People took on roles—mountain dwellers, sea dwellers, spirit beasts. The only way to ‘level up’ was by contributing to the story itself.” (P13)}
\end{quote}

Participants collectively world-built a symbolic narrative that connected diverse activities into a shared imaginary.
The Social Layer’s open architecture supported these evolving fictions, letting stories, rituals, and identities emerge from within.

\subsubsection{Mini Case: \textit{Counting Rice}}

\begin{figure}
    \centering
    \includegraphics[width=1\linewidth]{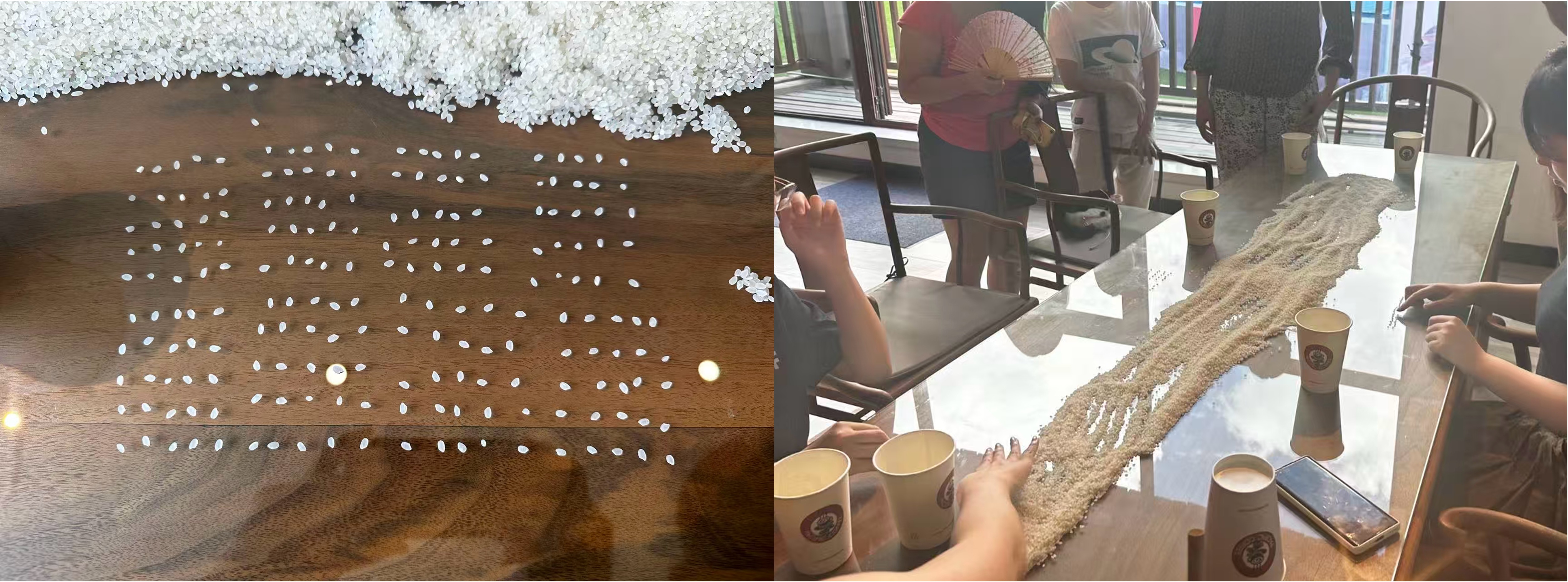}
    \caption{Improvisational Event: Counting Rice}
    \label{fig:rice}
\end{figure}

A defining micro-level example of serendipity in action was the event \textit{Counting Rice}, spontaneously initiated by a participant (\textit{P1}) during ShanHaiWoo 2023(Fig. \ref{fig:rice}).

The event occurred in a small, sunlit office space on the Shanhaiwoo residency site. \textit{P1} placed a event notice on Social Layer interface allowed participants to see the event immediately, RSVP, and join without any formal permission or approval. Within minutes, eight participants gathered, each approaching the task differently—some meticulously arranging grains into strict grids, others experimenting with diagonal patterns or playful abstractions.

Participants described a mixture of curiosity, amusement, and focused engagement:

\begin{quote}
\textit{“I didn’t know why I was doing it, but it felt like a small rebellion against usual productivity. Each grain was like a tiny decision, and together, we created a collective rhythm.” (P8)}
\end{quote}

\begin{quote}
\textit{“It was utterly meaningless, yet deeply absorbing. The room felt like a shared mental playground, where everyone’s small actions influenced the emergent whole.” (P6)}
\end{quote}

Participants later reflected that the event would likely never occur in a traditional structured setting:

\begin{quote}
\textit{“It felt useless but fun. I cannot imagine it happening anywhere else. The fact that I could just start something—without asking anyone—made me feel I could shape this place.” (P2)}
\end{quote}

\begin{quote}
\textit{“These micro-rituals made the residency alive. You could sense things forming—like small gravitational fields—around spontaneous gatherings.” (P4)}
\end{quote}

Through \textit{Counting Rice}, we see how low-friction tools, visible cues, and an open social architecture coalesced to create an emergent, playful, and socially meaningful experience. The event exemplifies how infrastructuring conditions can transform trivial or absurd actions into rich, serendipitous interactions that enhance trust, collaboration, and creative exploration.

\subsection{Infrastructuring Serendipitious Co-living}
Across these deployments, Social Layer’s infrastructural affordances—visible participation cues, frictionless event creation, and flexible boundary modules—enabled serendipity not by prescribing it, but by sustaining conditions for its emergence. In this sense, the system did not ‘cause’ spontaneous encounters, but made them infrastructurally possible at community scale. The following section synthesizes these insights as design patterns for future improvisational infrastructures.

\section{Discussion}

% We propose scaffolded spontaneity as a strong concept—an intermediate‑level design idea that can guide the design of socio‑technical systems beyond this specific case.  ￼

% Scaffolded spontaneity names a particular balance between:
% 	•	Scaffolding: infrastructural supports that make it easy to invite, find, and remember events (shared schedules, low‑friction hosting, visibility of attendance); and
% 	•	Spontaneity: room for unplanned, playful, and emergent activity that may appear “useless” in conventional terms but is central to serendipitous connection.

% In Social Layer, this balance is achieved by:
% 	•	Prioritising event‑hosting as a first‑class action, signalling to participants that they are authorised to create;
% 	•	Maintaining social translucence—making activity visible without overly personalising or prescribing paths;  ￼
% 	•	Treating identity and boundaries as configurable, allowing communities to tune openness and trust; and
% 	•	Providing a memory layer that preserves traces without fixing interpretations.

% Scaffolded spontaneity differs from generic “serendipity” in that it foregrounds designable conditions rather than the experience itself. It also connects to temporary urbanism by suggesting how ephemeral infrastructures can still support meaningful, world‑building practices over short periods.  ￼

% As a strong concept, scaffolded spontaneity can be contested, refined, and applied to other settings: corporate retreats, university intensives, civic prototyping labs, or digital‑only communities.

\subsection{Scaffolded Spontaneity: Improvisational Infrastructures at the Edge of Chaos}

We articulate scaffolded spontaneity as a strong concept to describe how infrastructures can be designed as sets of affordances that invite improvisation rather than prescribe programmes. In our account, scaffolded spontaneity refers to infrastructural conditions that make it easy, safe, and legible for participants to initiate and recombine activities, while keeping the collective near the “edge of chaos” rather than in full order or entropy \cite{hollandHiddenOrderHow2003, kauffmanOriginsOrderSelfOrganization1993}.

Social Layer operationalises this concept through concrete affordances: a prominent “Create Event” button with ultra-minimal schema (“what, where, when”), a shared schedule and map view that render all activities co-visible in time and space, lightweight RSVP and presence cues, and flexible boundary and role settings that treat rules as editable rather than hard-coded.  These structures do not dictate content; they signal how to start something and where it might fit. Participants explicitly linked these affordances to their sense of agency: Social Layer “made it effortless to start something—it removed permission, roles, and the need for approval,” and “amplified individual agency—turning a personal spark into collective action.” 

Our macro-level findings show how these affordances scaled. Across five deployments, 69.7\% of 1,586 events were self-organized, with Shanhaiwoo and Gathering of Tribe reaching 95–98\% member-initiated events within highly compressed timeframes.  Micro-cases such as Counting Rice—an ostensibly "useless" but "artsy" activity that quickly gathered eight participants into an absorbing shared experiment—demonstrate how low-friction initiation, shared visibility, and open invitations can turn trivial sparks into collective improvisation and world-building.

Design-wise, scaffolded spontaneity foregrounds calibrated friction as a key lever. Requiring every event to have a host, time, and place, and to appear on a common schedule, introduced “just enough” friction to prevent collapse into noise, while keeping initiation near zero-cost.  Treating user-initiated actions as first-class interactions, instrumenting shared rhythms (meals, unconference slots, rituals), and allowing communities to tune rules over time together compose a repertoire of affordances through which improvisational infrastructures can be designed, tested, and iterated in pop-up cities. 

\subsection{Serendipitous Criticality: Weak-Tie Bridging in Temporary Co-living}

Our findings reinforce social theory about the strength of weak ties \cite{granovetterStrengthWeakTies1973} and illustrate how technology can facilitate weak-tie interaction in temporary physical communities. Traditional community design often focuses on strengthening bonds among members (bonding capital), but bridging capital – connecting across diverse groups – is equally vital for innovation and learning \cite{burtStructuralHolesGood2004}. Social Layer’s design patterns (open invitations, visibility of others’ interests, prompt to “see what’s happening nearby”) intentionally surface opportunities for people to step outside their usual circles. We observed that this led to many cross-pollinating encounters that participants valued. For design researchers, this suggests designing “bridge prompts” – features that gently encourage interaction between otherwise unconnected subgroups. Examples might include highlighting events that draw mixed attendance, or suggesting collaborative activities that span disciplines. The key is that these are not personalized recommendations (“You and X should meet”), but rather environmental nudges (like public postings of activities) that rely on human initiative to bridge. This respects individual agency and avoids over-curation, aligning with philosophy that serendipity cannot be forced, only invited \cite{andreDiscoveryNeverChance2009, bjornebornThreeKeyAffordances2017}. In sum, surfacing weak-tie opportunities should be a design goal for systems aiming to enhance community serendipity. The benefits are improved creativity, faster dissemination of ideas, and a more inclusive social mesh, as newcomers or peripheral members find entry points to engage.

\subsection{Limitations and Future Work}

While our findings are encouraging about the role of digital infrastructure in fostering serendipitous co-living, there are important limitations and open questions. First, our data relied on cases that were all somewhat aligned in ethos – these were communities predisposed to openness and tech friendliness. We do not know how Social Layer would fare in a community with more resistant culture or where digital tools are viewed skeptically. Future work could deploy similar systems in more varied contexts, or even in corporate innovation workshops, to test generality.

Second, we did not conduct systematic post-hoc interviews in this study (to avoid overburdening participants who already had reflection rituals). Our insights come from observations and voluntary feedback. A more structured evaluation (e.g., measuring social network change or surveying perceived serendipity) would strengthen the evidence. We plan to complement this with trace ethnography and interviews in upcoming deployments to assess long-term outcomes: e.g., did people form collaborations that persisted months later? How inclusive was the spontaneity – did everyone feel equally empowered to host, or did subtle hierarchies remain (like language barriers, etc.)? These questions touch on inclusion and well-being, which are vital if such infrastructures are to be heralded as socially beneficial.

A tension worth exploring is scaling and overload. In Edge Esmeralda (580 people), some reported feeling overwhelmed by the sheer number of events (hundreds in 4 weeks). Serendipity can have a paradox: too many choices might lead to decision fatigue or fragmentation of community (everyone in their niche micro-events). Designing interfaces that help surface relevant serendipity without overwhelming is an open challenge. We consciously avoided heavy recommendation, but maybe gentle personalization could help in larger settings (while still preserving transparency). There’s space for research on how to balance discovery and focus, perhaps through better visualization or community moderation (Edge Esmeralda introduced a daily digest meeting to highlight certain events).

\section{Conclusion}
Pop‑up cities are a distinct socio‑technical form: longer than conferences, shorter than degrees; more structured than everyday urban anonymity yet less scripted than academic programs. By analyzing the design and deployment of Social Layer across multiple communities, we articulated “scaffolded spontaneity” and documented infrastructuring moves that enable serendipity while maintaining trust, rhythms, and care. Our account bridges CSCW/STS notions of infrastructuring and boundary objects with HCI concerns about social translucence and with urban studies on temporary urbanism. Future work will combine in‑depth interviews with trace ethnography to evaluate how these infrastructures shape inclusion, well‑being, and long‑term collaboration—so that temporary cities can be both playful and profound.

\bibliographystyle{ACM-Reference-Format}
\bibliography{reference}

\end{document}